\documentclass[aps,reprint]{revtex4-1}
\pdfoutput=1

\usepackage{braket}
\usepackage{amsmath}
\usepackage{amsthm}
\usepackage{amssymb}
\usepackage{mathrsfs}
\usepackage{mathtools}
\usepackage{color}
\usepackage{comment}
\usepackage{multirow}

\makeatletter

\newcommand{\Rmnum}[1]{\expandafter\@slowromancap\romannumeral #1@}
\makeatother

\begin{document}

\title{Interplay between spontaneous decay rates and Lamb shifts in open photonic systems}

\author{Emmanuel Lassalle}
\email[]{emmanuel.lassalle@fresnel.fr}
\affiliation{Aix Marseille Univ, CNRS, Centrale Marseille, Institut Fresnel, Marseille, France}

\author{Nicolas Bonod}
\affiliation{Aix Marseille Univ, CNRS, Centrale Marseille, Institut Fresnel, Marseille, France}

\author{Thomas Durt}
\affiliation{Aix Marseille Univ, CNRS, Centrale Marseille, Institut
  Fresnel, Marseille, France}

\author{Brian Stout}
\email[]{brian.stout@fresnel.fr}
\affiliation{Aix Marseille Univ, CNRS, Centrale Marseille, Institut Fresnel, Marseille, France}

\date{\today}

\begin{abstract}
In this letter, we describe the modified decay rate and photonic Lamb (frequency) shift of quantum emitters in terms of the
resonant states of a neighboring photonic resonator. 
This description illustrates a fundamental distinction in the behaviors 
of closed (conservative) and open (dissipative) systems:
the Lamb shift is bounded
by the emission linewidth in closed systems while it overcomes
this limit in open systems. 
\end{abstract}

\maketitle

\section{Introduction}

The coupling between quantum emitters (QE) and resonant photonic nanostructures
is at the heart of nanophotonics \cite{koenderink2017single}. The
resonances of a given structure characterize its optical response to
an excitation electromagnetic (EM) field, and can be of different
nature: Mie resonances in dielectric
structures \cite{decker2016resonant}, or surface plasmons in
metallic ones \cite{marquier2017revisiting}. A powerful tool
to describe both types of resonances is the use of resonant states, also called quasi-normal modes (QNMs), which are the natural modes of the
photonic system. A remarkable advantage of this framework is that it
allows to generalize the usual cavity-quantum electrodynamics (cQED) figures of merit characterizing the interaction between
a dipole source and the resonance of a cavity, such as the
quality factor $Q$ or mode volume $V$, to the case of open and/or
absorbing systems (and also taking into
account material dispersion) that are almost
always found in nanophotonics
\cite{kristensen2012generalized,sauvan2013theory,muljarov2016exact}. For
  instance, the use of QNMs has been proposed to express the
spontaneous decay rate of a QE coupled to an open photonic resonator in \cite{sauvan2013theory,muljarov2016exact}.

In this work, we express the photonic Lamb shift (\emph{i.e.} the
  shift of the emission frequency due to the neighboring environment)
  in terms of the QNMs. We start from the general quantum optics results,
valid in the weak-coupling regime, that relate
the spontaneous decay rate and photonic Lamb shift of a two-level QE to the Green tensor.
We then make use of the resonant states of the photonic
system to which the QE couples to expand the Green tensor and derive
analytic expressions for the decay rate and Lamb shift involving the
following figures of merit: quality factor, mode volume and Purcell factor. The derived
expressions reveal an interplay between the decay rate and the Lamb
shift, and in the single-resonance limit, we show that: (i) for
conservative (Hermitian) systems, \emph{i.e.} closed and
non-absorbing, or for systems with small
losses characterized by high quality factors,
the Lamb shift always lies within the emission linewidth (equal to the decay rate), whereas (ii) for
dissipative, \emph{i.e.} open and/or absorbing, systems (non-Hermitian), the Lamb shift can go beyond this fundamental
limit (see \cite{van2012spontaneous} with references therein and \cite{lassalle2017lamb}).

\section{Lamb shift and decay rate interplay}

In the presence of a neighboring photonic structure, a two-level QE will experience a new
decay rate $\gamma^*$ and
emission frequency $\omega^*$ compared to the case of free-space denoted by $\gamma_0$ and $\omega_0$, respectively.
Using the macroscopic quantum optics results for a two-level atom
coupled to a general dispersive and absorbing
medium, one can write the environment-modified decay rate $\gamma^*$ and photonic
Lamb shift $\Delta\omega$ defined by $\Delta\omega\equiv\omega^*-\omega_0$
in terms of the Green tensor $\overset\leftrightarrow{\bold{G}}$
(defined in Appendix~\ref{app:green})
that fully contains all the
properties of the EM environment of the emitter (we assume a constant relative
permeability $\mu=1$) \cite{dung2000spontaneous,dung2001decay}
\begin{equation}
\frac{\gamma^*}{\gamma_0}  = 1+  \frac{6\pi
  c}{\omega_0}\times \bold{u}_p\cdot
\text{Im}(\overset\leftrightarrow{\bold{G}}_s(\bold{r}_0,\bold{r}_0,\omega_0))\cdot\bold{u}_p \;,
\label{eq:decay_norm}
\end{equation}
\begin{equation}
\frac{\Delta\omega}{\gamma_0} = - \frac{3\pi
  c}{\omega_0}\times \bold{u}_p\cdot
\text{Re}(\overset\leftrightarrow{\bold{G}}_s(\bold{r}_0,\bold{r}_0,\omega_0))\cdot\bold{u}_p \;,
\label{eq:lamb_norm}
\end{equation}
where $c$ is the speed of
light in vacuum, $\bold{r}_0$ the atom position and $\bold{u}_p$ the unit vector in the direction of its
electric dipole moment: $\bold{p}_0=p_0\bold{u}_p$. Note that in these expressions, the total
Green tensor $\overset\leftrightarrow{\bold{G}}$ has been decomposed into the "free-space"
part $\overset\leftrightarrow{\bold{G}}_0$ (\emph{i.e.} the Green tensor in the
absence of the neighboring structure)
and a "scattered" part
$\overset\leftrightarrow{\bold{G}}_s$ (which defines the contribution
of the photonic system) as:
$\overset\leftrightarrow{\bold{G}}=\overset\leftrightarrow{\bold{G}}_0+\overset\leftrightarrow{\bold{G}}_s$,
and the quantities have been normalized by the free-space decay rate 
$\gamma_0=2\omega_0^2/(\epsilon_0 \hbar c^2)|\bold{p}_0|^2\bold{u}_p\cdot
\text{Im}(\overset\leftrightarrow{\bold{G}}_0(\bold{r}_0,\bold{r}_0,\omega_0))\cdot\bold{u}_p$
where $\bold{u}_p\cdot
\text{Im}(\overset\leftrightarrow{\bold{G}}_0(\bold{r}_0,\bold{r}_0,\omega_0))\cdot\bold{u}_p=\omega_0/6\pi
c$.
Moreover, as far as the Lamb shift is concerned,
the integral part over all
frequencies (see \cite{dung2001decay}) has been omitted. 
Eqs.~(\ref{eq:decay_norm}) and (\ref{eq:lamb_norm})
obtained within a two-level system model are the
same as the ones of the decay rate and radiative frequency-shift of
a classical electric dipole normalized by the
classical decay rate in free space \cite{novotny2012principles}. A
more complete treatment of the Lamb shift for real multilevel atoms can be found in \cite{PhysRevA.32.2030}.

We assume that the scattered part of the Green tensor
$\overset\leftrightarrow{\bold{G}}_s$ can be expanded in terms
of the resonant states of the photonic system, and
we use the spectral representation of the Green tensor \cite{sauvan2014modal,doost2014resonant,muljarov2016exact}
\begin{equation}\label{eq:green_expand}
\overset\leftrightarrow{\bold{G}}_s(\bold{r},\bold{r'},\omega) \simeq
c^2\sum_\alpha\frac{\bold{E}_\alpha(\bold{r})\otimes\bold{E}_\alpha(\bold{r'})}{2\omega(\omega_\alpha-\omega)}
\end{equation}
where $\bold{E}_\alpha$ are the
QNM fields normalized according to Muljarov \emph{et al.}
\cite{doost2014resonant,muljarov2016exact}, $\omega_\alpha \equiv
\omega_\alpha' + \text{i}\,\omega_\alpha''$ are the QNM complex
frequencies and $\otimes$ denotes the tensor product (definitions in
Appendix \ref{app:resonant_states}).
When plugging Eq.~(\ref{eq:green_expand}) in Eqs.~(\ref{eq:decay_norm})
and (\ref{eq:lamb_norm}), one immediately gets
\begin{equation}
\frac{\gamma^*}{\gamma_0} = 1 + \frac{3\pi\,
  c^3}{\omega_0^2}\,\sum_\alpha\text{Im}\left(\frac{1}{V_\alpha(\omega_\alpha-\omega_0)}\right)
\label{eqapp:decayenh}
\end{equation}
\begin{equation}
\frac{\Delta\omega}{\gamma_0}  = - \frac{3\pi\,c^3}{2\omega_0^2}\sum_\alpha\text{Re}\left(\frac{1}{V_\alpha(\omega_\alpha-\omega_0)}\right)
\label{eqapp:LBveff}
\end{equation}
where $V_\alpha$ is the mode volume of the QNM $\alpha$ defined as
\begin{equation}
V_\alpha \equiv \frac{1}{(\bold{u_p}\cdot\bold{E}_\alpha(\bold{r}_0))^2}
\label{eqapp:eff_vol}
\end{equation}
in which the QNM field $\bold{E}_\alpha$ is taken at the QE position $\bold{r}_0$.
This figure of merit characterizes the coupling between the QE and the
resonance $\alpha$ through the real part (the larger
$\text{Re}(1/V_\alpha)$, the better is the coupling) \cite{zambrana2015purcell}, and also
energy dissipations through the presence of an imaginary part (a
large $\text{Im}(1/V_\alpha)$ indicates important energy dissipations) \cite{sauvan2013theory}. 
We next introduce the Purcell factor $F_\alpha$, which corresponds to
the enhancement of the total
decay rate $\gamma^*$ in comparison to $\gamma_0$ due the the resonance
$\alpha$ and for a perfect
spectral match ($\omega_0=\omega_\alpha'$),
\begin{equation}
F_\alpha \equiv \frac{6\pi\,c^3}{\omega_\alpha'^3}Q_\alpha\text{Re}\left(
  1/V_\alpha\right)  \;,
\label{eq:purcell_factor}
\end{equation}
with the usual quality factor $Q_\alpha$ defined as $Q_\alpha \equiv -
\omega_\alpha'/(2\omega_\alpha'')$ ($\omega_\alpha'' < 0$
due to the convention used for the
Fourier transform ``$\,e^{-\text{i}\omega t}\,$''). Expressions~(\ref{eqapp:decayenh}) and
(\ref{eqapp:LBveff}) can then be recast in a form revealing an
interplay between Lamb shift and decay rate (see derivation in
Appendix~\ref{app:derivation1})
\begin{equation}
\frac{\gamma^*}{\gamma_0}  = 1+
\sum_\alpha \left \{ \frac{\gamma_\alpha^H}{\gamma_0} -
  2\frac{\Delta\omega_\alpha^H}{\gamma_0}\frac{\text{Im}(1/V_\alpha)}{\text{Re}(1/V_\alpha)}
\right \}
\label{eq:decay:univ}
\end{equation}
\begin{equation}
\frac{\Delta\omega}{\gamma_0}  = \sum_\alpha
\left \{ \frac{\Delta\omega_\alpha^H}{\gamma_0} +
\frac{1}{2}\frac{\gamma_\alpha^H}{\gamma_0}\frac{\text{Im}(1/V_\alpha)}{\text{Re}(1/V_\alpha)}
\right \}
\label{eq:lamb:univ}
\end{equation}
where the expressions of $\gamma_\alpha^H/\gamma_0$ and $\Delta\omega_\alpha^H/\gamma_0$
are 
\begin{equation}
\frac{\gamma_\alpha^H}{\gamma_0} = F_\alpha \left(
  \frac{\omega_\alpha'}{\omega_0} \right)^2
\frac{\omega_\alpha''^2}{(\omega_\alpha' - \omega_0)^2 +
  \omega_\alpha''^2} 
\label{eq:gamma_hermitian}
\end{equation}
\begin{equation}
\frac{\Delta\omega_\alpha^H}{\gamma_0} = F_\alpha \left(
  \frac{\omega_\alpha'}{\omega_0} \right)^2 \frac{\omega_\alpha''}{2} \frac{\omega_\alpha'-\omega_0}{(\omega_\alpha' - \omega_0)^2 +
  \omega_\alpha''^2}\; .
\label{eq:lb_hermitian}
\end{equation}
The superscript $H$ indicates ``Hermitian'', because for
conservative systems, and more realistically for systems with small losses,
$\text{Im}(V_\alpha)\simeq 0$ and one recovers the sum of
Lorentzians which is phenomenologically used for high-Q
cavities \cite{sauvan2013theory}: $\gamma^*/\gamma_0  = 1+\sum_\alpha
\gamma_\alpha^H/\gamma_0$. In contrast, for
dissipative systems characterized by $\text{Im}(V_\alpha)\ne 0$ \cite{sauvan2013theory},
Eqs.~(\ref{eq:decay:univ}) and (\ref{eq:lamb:univ}) reveal an interplay
between the ``Hermitian'' decay rates $\gamma_\alpha^H$ and Lamb
shifts $\Delta\omega_\alpha^H$. This constitutes our first
result.


\section{Example} 

\begin{figure}
  \centering
   \includegraphics[width=\linewidth]{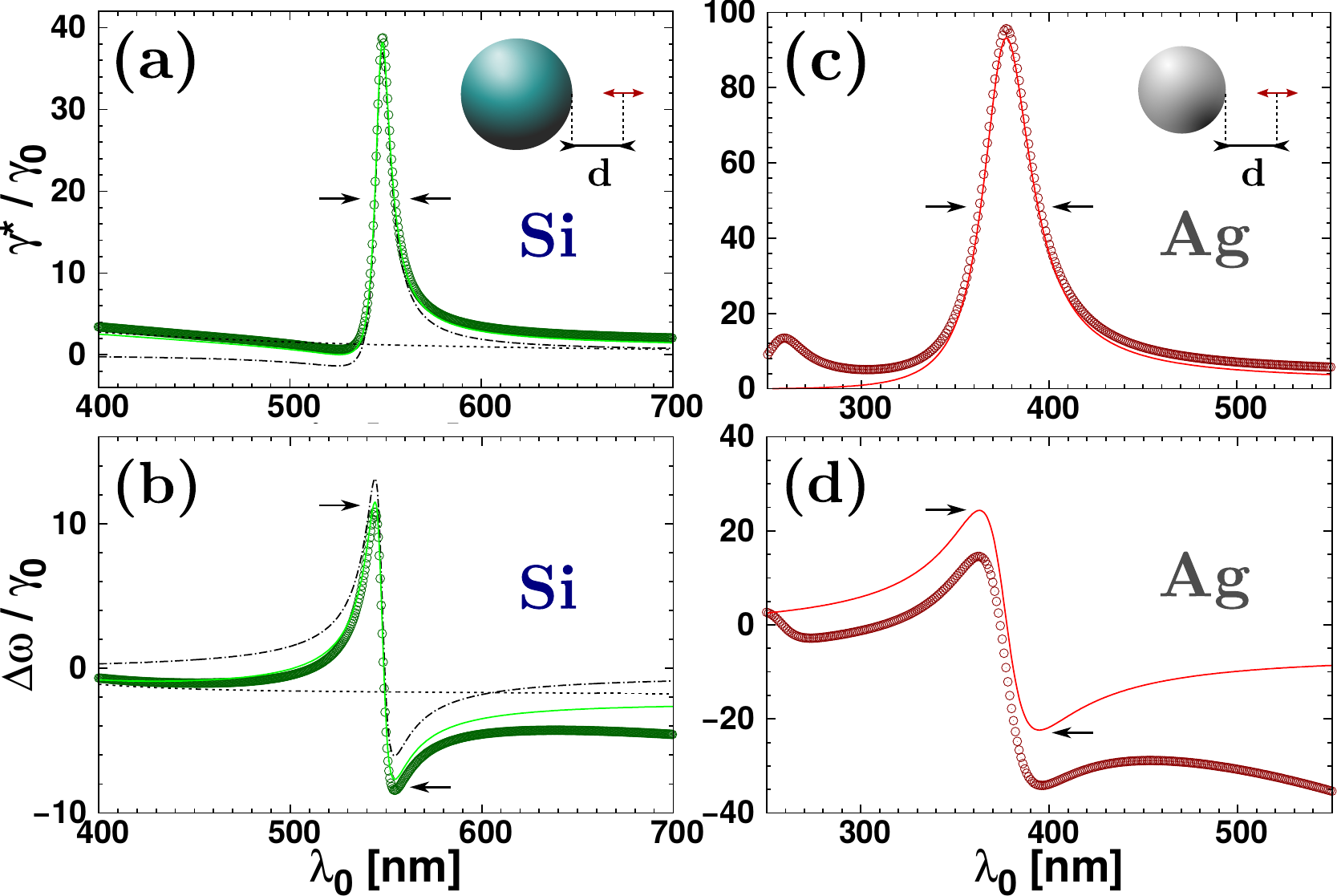}
  \caption{Comparison between QNM calculations using
    Eqs.~(\ref{eq:decay:univ}) and
(\ref{eq:lamb:univ}) (lines) and exact calculations using Mie
theory (circles)
  of the decay
  rate $\gamma^*$ and Lamb shift $\Delta\omega$ (normalized by $\gamma_0$) as a function of the emitter transition
  wavelength $\lambda_0=2\pi c/\omega_0$, for two
  configurations: (a-b) silicon (Si) and (c-d) silver (Ag) nanospheres of
  radii $a=120\,\text{nm}$ and $a=20\,\text{nm}$, respectively. In all
  cases, the emitter (red arrow) is radially
oriented and located at a distance $d=10\,\text{nm}$ from the sphere. For Si [Ag], only the electric quadrupolar [dipolar]
  contribution to the decay rate (a) [(c)] and Lamb
  shift (b) [(d)] is shown.} 
  \label{fig:dielectric}
\end{figure}

As an example, let us apply Eqs.~(\ref{eq:decay:univ}) and
(\ref{eq:lamb:univ}) to two situations of a QE coupled to an open
photonic system: (i) dielectric silicon (Si) nanosphere with no
absorption (and no dispersion) and (ii) plasmonic silver (Ag)
nanosphere with absorption (and dispersion) (see insets in Fig.~\ref{fig:dielectric} (a-c)). 
For spherical
resonators, the QNM fields are the multipolar fields, labeled by four numbers
$\{q,n,m,l\}$ where $q$ labels a magnetic ($q=1$) or an electric
($q=2$) mode, $n=1,2,...,\infty$ is the multipolar order, $m=-n,...,n$
is the orbital (or azimutal) number, and $l$ numerates the
different QNM complex frequencies $\omega_{q,n,m,l}$ found for a fixed
combination of $\{q,n,m\}$, which are the poles of the
Mie coefficients \cite{stout2011multipole,zambrana2015purcell,lassalle2017lamb}.
Therefore, for spherical resonators the sums over $\alpha$ in Eqs.~(\ref{eq:decay:univ}) and
(\ref{eq:lamb:univ}) become:
  $\sum_\alpha\rightarrow \sum_{q,n,m,l}$. Moreover, for a given set of $\{q,n,l\}$, the QNMs with a different number $m$ are
degenerate (\emph{i.e.} have the same complex frequency
$\omega_{q,n,m,l}$), and the sum can be
recast in the form $\sum_{q,n,l}$ with ``effective'' mode volumes
defined as $1/V_{q,n,l}\equiv\sum_m1/V_{q,n,m,l}$ (see Appendix E in \cite{muljarov2016exact} and also \cite{lassalle2017lamb}).

In the following, the mode volumes given by
Eq.~(\ref{eqapp:eff_vol}), and appearing in the QNM formulas Eqs.~(\ref{eq:decay:univ}) and
(\ref{eq:lamb:univ}), are computed using the analytical expressions
of the QNM fields $\bold{E}_\alpha$ derived for a spherical resonator
shape in \cite{doost2014resonant} for non-dispersive materials
(\emph{i.e.} with a constant permittivity) and in
\cite{muljarov2016exact} for dispersive materials (\emph{i.e.} with a
permittivity that depends on the frequency $\omega$). The QNM
complex frequencies are found by solving a transcendental equation (giving the poles of the Mie
coefficients) with the FindRoot function of Mathematica, and where
we use an analytic continuation of the permittivity in the complex plane in the case of dispersive materials.

\begin{table}[bp]
\centering
\begin{tabular}{|c|c|c|c|}
\hline
$$ & $\alpha = (n,l)$ & $\lambda_\alpha \, (\text{nm})$ & $V_\alpha \, (\text{nm}^3)$ \\
\hline
\multirow{2}{*}{Si} & $(2,1)$ & $547.3 + \mathrm{i}4.7$ & $(17.573 -\mathrm{i}6.974)\cdot 10^{6}$ \\
 & $(2,2)$ & $329.7 + \mathrm{i}106.3$ & $(1.222 + \mathrm{i}1.063)\cdot 10^{6}$ \\
\hline
  \multirow{1}{*}{Ag} & $(1,1)$ & $375.6 + \mathrm{i}15.5$ & $(0.525 - \mathrm{i}0.023)\cdot 10^{6}$ \\
\hline
\end{tabular}
\caption{QNMs complex wavelengths $\lambda_\alpha$ and mode volumes $V_\alpha$ appearing in Eqs.~(\ref{eq:decay:univ}) and
(\ref{eq:lamb:univ}) and used to obtain the results of Fig.~\ref{fig:dielectric}.}
  \label{tab:complex_frequencies}
\end{table}

For the calculations, we consider an electric dipole emitter radially
oriented (and therefore only coupled to the electric modes $q=2$
\cite{rolly2012promoting}) and located at a distance $d=10\,\text{nm}$ from the
sphere. The Si nanosphere (dielectric permittivity $\varepsilon =
16$) has a radius of $a=120\,\text{nm}$, exhibiting a
dominant electric quadrupolar resonance at $547$ nm,
and the Ag nanosphere (Drude-Lorentz model for the dielectric permittivity taken
from \cite{rakic1998optical}) has a radius of $a=20\,\text{nm}$, exhibiting a dominant
electric dipolar resonance at $375$ nm. 
For the Si configuration,
we show in Figs.~\ref{fig:dielectric} (a-b)
the electric quadrupolar contribution ($n=2$) to
the decay rate $\gamma^*$ and Lamb shift $\Delta\omega$ as a function of the emitter
transition wavelength $\lambda_0=2\pi c/\omega_0$, calculated from the QNM formulas~(\ref{eq:decay:univ}) and
(\ref{eq:lamb:univ}) (solid green line), and compared with the exact Mie theory
(green dots). We find several QNMs associated with this quadrupolar
resonance, and by using the two dominant QNMs
(whose mode volumes and complex
wavelengths defined as
$\lambda_\alpha\equiv 2\pi c/\omega_\alpha$ are given in
Table~\ref{tab:complex_frequencies}), the QNM
formulas work very well, with a better result for the decay
rate than for the Lamb shift for which one can see a certain discrepancy at high
wavelengths. The two individual contributions of the QNMs used in the expansion are also shown (dashed black lines in (a-b)).

For the Ag
configuration we show in Figs.~\ref{fig:dielectric}
(c-d) the dominant dipolar contribution ($n=1$) to the decay rate $\gamma^*$ and Lamb shift $\Delta\omega$ as a function of the emitter
transition wavelength $\lambda_0$, calculated from Eqs.~(\ref{eq:decay:univ}) and
(\ref{eq:lamb:univ}) (solid red lines) and compared with the Mie
theory (red dots). We only find a single QNM associated with this dipolar resonance (whose mode volume and complex
wavelength are given in Table~\ref{tab:complex_frequencies}). The agreements are
quite good for the decay rate, but one can
see certain discrepancies for the Lamb shift (the resonance around
$250\,\text{nm}$ is a spurious resonance peculiar to the model of
permittivity used \cite{hao2007efficient}). 
These Lamb shift discrepancies (more important in the metallic case) appear to be related to
omitted non-resonant contributions (see Eq.~(4) in \cite{grigoriev2013optimization} and
Eq.~(16) in \cite{zhan2014theory}), which impact more the
Lamb shift than the decay rate in the near field.
Finally,
let us emphasize the presence of an imaginary part in the mode volumes
displayed in Table~\ref{tab:complex_frequencies}. In the dielectric case,
the imaginary part characterizes the radiative
losses and in the plasmonic case,
it characterizes both radiative and absorption losses.

\section{Maximum Lamb shifts in the single-resonance approximation}

From here on, we work under the assumption that the QE couples to a single resonance $\alpha$. First, we revisit the
case of conservative or low-loss systems for which $\text{Im}(V_\alpha)\simeq 0$.
In this case,
Eqs.~(\ref{eq:decay:univ}) and (\ref{eq:lamb:univ}) become
$\gamma^*/\gamma_0 = 1+
\gamma_\alpha^H/\gamma_0
\text{ and }
\Delta\omega/\gamma_0 = 
\Delta\omega_\alpha^H/\gamma_0$,
and we can see that the decay rate
$\gamma_\alpha^H$ and the Lamb
shift $\Delta\omega_\alpha^H$ are dissociated and there is no interplay.
We want to assess the maximum frequency shift
$\Delta\omega_{\text{max}}$, that occurs when the
QE natural frequency $\omega_0$ is detuned by $\pm \omega_\alpha''$ compared to the QNM resonance
frequency $\omega_\alpha'$. At these particular frequencies
$\omega_0  = \omega_\alpha' \mp \omega_\alpha''$, the decay rate and
Lamb shift (pointed out with arrows in
Fig.~\ref{fig:dielectric}) are (see Appendix~\ref{app:derivation2})
\begin{equation}
\frac{\gamma^*}{\gamma_0} = 1+\frac{1}{2}F_\alpha + O( Q_\alpha^{-1})
\label{eq:gamma_-_h}
\end{equation}
\begin{equation}
\frac{\Delta\omega_{\text{max}}}{\gamma_0}=
\pm \frac{1}{4}F_\alpha +O( Q_\alpha^{-1})
\label{eq:lamb_-_h}
\end{equation}
($\Delta\omega_{\text{max}} \simeq +1/4F_\alpha$ when $\omega_0  = \omega_\alpha' - \omega_\alpha''$
and $\Delta\omega_{\text{max}} \simeq -1/4F_\alpha $ when $\omega_0  = \omega_\alpha' + \omega_\alpha''$).
We retrieve in this ideal case the expressions for the maximum frequency
shift that were derived in \cite{ching1987dielectric}, Eq.~(35), where they considered a
two-level atom inside a cavity whose resonance was
phenomenologically described by a Lorentzian. For large decay rate
enhancements $\gamma^*\gg \gamma_0$, the first term in
the right hand side of Eq.~(\ref{eq:gamma_-_h}) can be omitted and we finally end up with the following
relations for the maximum photonic Lamb shift
\begin{equation}
\Delta\omega_{\text{max}} = \pm \frac{\gamma^*}{2}\; .
\label{eq:heisenberg1}
\end{equation}

Before commenting this result, let us first recall that in the
weak-coupling regime, the emitted-light spectrum of the QE has a Lorentzian line
shape \cite{dung2000spontaneous},
and one usually takes the full width at half
maximum (FWHM) $\hbar\gamma^*$ as a measure of the
energy spread $\delta E$, called energy level width or emission linewidth. This leads to the relation
between the energy level width and the lifetime of the excited state
(defined as $\tau \equiv 1/\gamma^*$): $\delta E \, \tau = \hbar$, which can be
seen as a time-energy uncertainty relation (see \emph{e.g.} \cite{dodonov2015energy}).
Thus --- and this is our second result --- for conservative systems,
or systems with weak
energy dissipations, and in the single-resonance
approximation, Eq.~(\ref{eq:heisenberg1}) shows that the photonic Lamb shift always lies
within the emission linewidth. As already pointed out in
\cite{ching1987dielectric}, this makes it difficult to observe as a shift of the spectral
line.

\begin{figure}[htbp]
  \centering
  \includegraphics[scale=0.85]{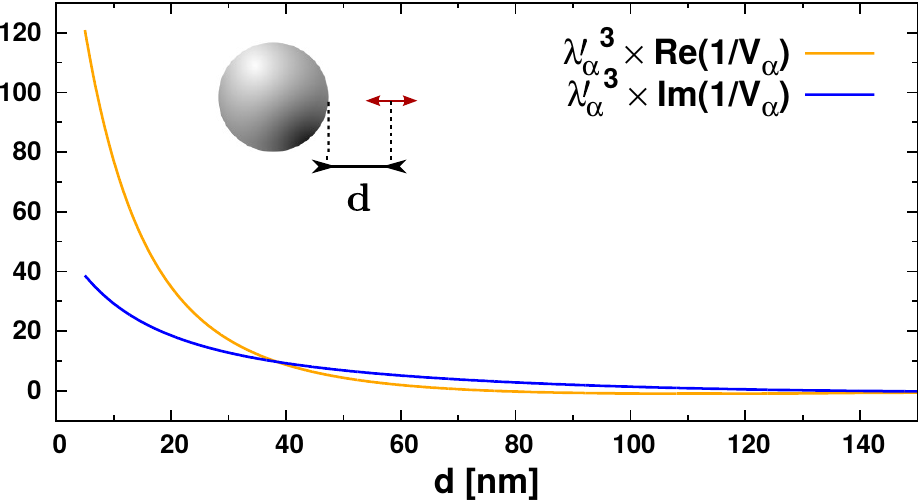}
  \caption{Inverse of the mode
    volume $1/V_\alpha$ of the dipolar QNM of a silver nanosphere
    (radius $a=50\,\text{nm}$), as
    a function of the distance $d$ (real part in orange and
    imaginary part in blue), for an emitter radially oriented (red arrow). Note that $1/V_\alpha$
    has been multiplied by the cube of the QNM resonance wavelength $\lambda'_\alpha=411.6\,\text{nm}$.}
  \label{fig:effective_volume}
\end{figure}

Let us now turn to the case of dissipative systems, for which
$\text{Im}(V_\alpha)\ne 0$. 
In this case, the decay rate and Lamb
shift are described by Eqs.~(\ref{eq:decay:univ}) and (\ref{eq:lamb:univ}), respectively.
At the frequencies $\omega_0  = \omega_\alpha' \mp \omega_\alpha''$, these expressions
reduce to (see Appendix~\ref{app:derivation2})
\begin{equation}
\frac{\gamma^*}{\gamma_0} = 1+\frac{1}{2}F_\alpha \left[ 1 \mp
\frac{\text{Im}(1/V_\alpha)}{\text{Re}(1/V_\alpha)} \right]+ O( Q_\alpha^{-1})
\label{eq:gamma_+}
\end{equation}
\begin{equation}
\frac{\Delta\omega_{\text{max}}}{\gamma_0} = \pm \frac{1}{4}F_\alpha \left[ 1 \pm
\frac{\text{Im}(1/V_\alpha)}{\text{Re}(1/V_\alpha)} \right]+ O( Q_\alpha^{-1})
\label{eq:lamb_+}
\end{equation}
(when $\omega_0  = \omega_\alpha' - \omega_\alpha''$ one must take the
upper sign and when $\omega_0  = \omega_\alpha' + \omega_\alpha''$ one
must take the lower sign).
For large decay rate enhancements $\gamma^*\gg \gamma_0$, the first
term in the right hand side of (\ref{eq:gamma_+}) can be neglected
and we get the following relation
between the maximum Lamb shift and decay rate
\begin{equation}
\Delta\omega_{\text{max}} = \pm\frac{\left[\text{Re}(1/V_\alpha) \pm \text{Im}(1/V_\alpha) \right]}{\left[\text{Re}(1/V_\alpha) \mp \text{Im}(1/V_\alpha)
\right]}\frac{\gamma^*}{2}\; .
\label{eq:relation2}
\end{equation}
In sharp contrast with Eq.~(\ref{eq:heisenberg1}) valid for conservative
or high-Q systems,
Eq.~(\ref{eq:relation2}) shows that for dissipative systems, the Lamb shift is not bounded
by the emission linewidth, and can go beyond this limit. This is our
third result.

To illustrate this fundamental distinction in the behavior of
conservative and dissipative systems, we consider in the following
a QE radially oriented and coupled to the plasmonic dipolar resonance of a silver
nanoparticle of radius $a=50\,\text{nm}$ (see inset in
Fig.~\ref{fig:effective_volume}). The complex wavelength
$\lambda_\alpha=\lambda_\alpha'+\mathrm{i}\lambda_\alpha''$ of the
dipolar QNM is
calculated to be $\lambda_\alpha = 411.6 + \text{i}50.8$ nm, which
gives a quality factor
$Q_\alpha=\lambda'_\alpha/(2\lambda''_\alpha)=4$. The QE
transition wavelength $\lambda_0=2\pi c/\omega_0$ is assumed to be
$\lambda_0=372$ nm. This corresponds to the case
$\omega_0=\omega_\alpha'-\omega_\alpha''$ for which the Lamb shift $\Delta\omega_{\text{max}}$ is
maximum and positive and given by Eq.~(\ref{eq:lamb_+}) (taking the
positive sign), and the decay rate $\gamma^*$ is the one given by
Eq.~(\ref{eq:gamma_+}) (taking the
negative sign).
First, we
plot in Fig.~\ref{fig:effective_volume} (the inverse of) the mode
volume $V_\alpha$ of the dipolar QNM as a function
of the distance $d$ between the QE and the nanoparticle. One can see
that $\text{Re}(1/V_\alpha)$, which characterizes the coupling between the QE and the nanoparticle,
increases as $d$ decreases (orange curve), which is in accordance with the expectation
that the coupling increases as the QE
gets closer to the resonator. Moreover, one can see the presence of
energy dissipations through a non-negligible $\text{Im}(1/V_\alpha)$
(blue curve), which is expected
when considering the low quality factor of the resonance $Q_\alpha=4$.

\begin{figure}[htbp]
  \centering
  \includegraphics[width=\linewidth]{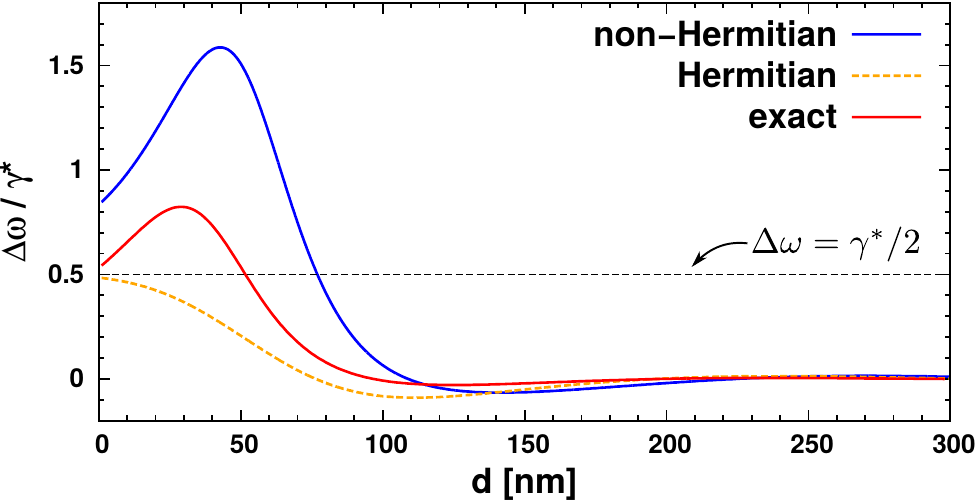}
  \caption{Ratio between the Lamb shift $\Delta\omega_{\text{max}}$ and the decay
    rate $\gamma*$ as a function of the distance $d$, calculated from Eqs.~(\ref{eq:gamma_+}) and
    (\ref{eq:lamb_+}) (in blue), from
    Eqs.~(\ref{eq:gamma_-_h}) and (\ref{eq:lamb_-_h}) (in orange), and from the Mie theory (in
    red), for the same configuration as in
Fig.~\ref{fig:effective_volume}. A guide-to-the-eye shows the limit $\Delta\omega=\gamma^*/2$.}
  \label{fig:decay_vs_LB}
\end{figure}

Accordingly, the decay rate $\gamma^*$ [Eq.~(\ref{eq:gamma_+})] and maximum Lamb shift
$\Delta\omega_{\text{max}}$ [Eq.~(\ref{eq:lamb_+})] will increase as $d$ decreases in a similar way as
$\text{Re}(1/V_\alpha)$ in Fig.~\ref{fig:effective_volume} (because the Purcell factor appearing in their expression is $F_\alpha
\propto \text{Re}(1/V_\alpha)$ [see Eq.~(\ref{eq:purcell_factor})]). More importantly, dissipations, through the presence
of $\text{Im}(1/V_\alpha)$ in Eqs.~(\ref{eq:gamma_+}) and (\ref{eq:lamb_+}), will weaken the decay rate (due to the negative
sign in Eq.~(\ref{eq:gamma_+})) and increase the Lamb shift (due to the positive sign in Eq.~(\ref{eq:lamb_+})), compared to the
conservative case where $\text{Im}(1/V_\alpha)=0$.
To see this effect, we plot in Fig.~\ref{fig:decay_vs_LB} the ratio
$\Delta\omega_\text{max}/\gamma^*$ as a function of the distance $d$,
for the dissipative case (blue curve) and the ideal conservative case (orange
curve). The limit $\Delta\omega=\gamma^*/2$ is also shown (dashed black line).
One can see that contrary to the conservative case where the Lamb shift
is bounded by $\gamma^*/2$, dissipations allow to
fulfill the condition $\Delta\omega>\gamma^*/2$. We compare
this result with the Mie calculations taking
into account only the electric dipolar ($n=1$) response of the nanoparticle
(red curve). Despite a decrease of the magnitude (that might be
explained by the non-resonant contributions discussed previously), the Mie calculations still show a Lamb shift that overcomes the limit of
conservative systems under a dipolar approximation.

\section{Conclusion}

To sum up, using a quasi-normal mode description, we derive general expressions for the
environment-modified decay rate and photonic Lamb shift, valid for open (dissipative) resonators. In the
single-resonance approximation, we consider the maximum level shift that can
be expected, and we show a remarkable difference between closed
(conservative) and
open (dissipative) systems: while for conservative systems, the Lamb
shift remains within the emission linewidth, it can
go beyond this fundamental range for dissipative systems.

\section{Appendix}

\subsection{Definition of the Green tensor}
\label{app:green}

The Green tensor
$\overset\leftrightarrow{\bold{G}}(\bold{r},\bold{r'},\omega)$ is
defined as the solution of the
classical Maxwell's equations with a $\delta$ function source term
\begin{equation}
\nabla\times\nabla\times\overset\leftrightarrow{\bold{G}}-\frac{\omega^2}{c^2}\varepsilon\,\overset\leftrightarrow{\bold{G}}
=\overset\leftrightarrow{\bold{I}}\delta(\bold{r}-\bold{r'})
\end{equation}
with the proper boundary
conditions. $\overset\leftrightarrow{\bold{I}}$ is the unit tensor and
$\varepsilon(\bold{r},\omega)$
is the relative permittivity. Note that we assume a constant relative
permeability $\mu=1$. The Green tensor defined by this equation has
the units: $[\overset\leftrightarrow{\bold{G}}]=\text{m}^{-1}$.

\subsection{Definition of the resonant states} 
\setcounter{equation}{0}
\renewcommand{\theequation}{B{\arabic{equation}}} 
\label{app:resonant_states}

The resonant states $\bold{E}_\alpha(\bold{r})$ of the photonic
system are defined as the solutions of the Maxwell's
equations in the absence of source 
\begin{equation}
\nabla\times\nabla\times\bold{E}_\alpha=\frac{\omega_\alpha^2}{c^2}\varepsilon(\bold{r},\omega)\bold{E}_\alpha
\end{equation}
where $\varepsilon$ is the relative permittivity of the
resonator and where a constant relative
permeability $\mu=1$. Moreover, these eigenmodes satisfy outgoing wave boundary conditions \cite{kristensen2012generalized,sauvan2013theory,doost2014resonant}. Because of the
boundary conditions, the eigenfrequencies $\omega_\alpha$ associated
to the eigenmodes $\bold{E}_\alpha$ are complex:
$\omega_\alpha=\omega'_\alpha+\text{i}\,\omega''_\alpha$, where $\omega_\alpha'' < 0$
due to the convention used for the Fourier transform ``$\,e^{-\text{i}\omega t}\,$''.
In this letter, we follow the normalization condition of
Doost \emph{et al.} \cite{doost2014resonant,muljarov2016exact} where the resonant modes
are normalized according to
\begin{multline}\label{app:norm}
1=\frac{1}{2}\int_V\,\bold{E}_\alpha\cdot\left[\frac{\partial(\omega\varepsilon)}{\partial(\omega)}+\varepsilon\right]\bold{E}_\alpha\,\mathrm{d}\bold{r}\\
+\frac{c^2}{2\omega_\alpha^2}\oint_{\partial
  V}\,\left[\bold{E}_\alpha\cdot \frac{\partial}{\partial
  s}(\bold{r}\cdot\nabla)\bold{E}_\alpha-(\bold{r}\cdot\nabla)\bold{E}_\alpha\cdot\frac{\partial
  \bold{E}_\alpha}{\partial
  s}\right]\mathrm{d}S\;.
\end{multline}
All the quantities that depend on $\omega$ are taken at $\omega=\omega_\alpha$. The first integral is taken over a volume $V$ enclosing the
photonic system, and the second integral is taken over a closed
surface $\partial V$ of the volume $V$, with the normal derivative
$\partial/\partial s =\bold{n}\cdot \bold{\nabla}$, $\bold{n}$ being
the outward unit vector normal to the surface. This normalization sets
the unity of the electric fields as: $[\bold{E}_\alpha]=\text{m}^{-3/2}$.

\subsection{Derivation of Eqs.~(\ref{eq:decay:univ}), (\ref{eq:lamb:univ}), (\ref{eq:gamma_hermitian})
and (\ref{eq:lb_hermitian})}
\label{app:derivation1}
Here we derive the interplay relations between the decay rate and Lamb
shift (\ref{eq:decay:univ}), (\ref{eq:lamb:univ}), (\ref{eq:gamma_hermitian})
and (\ref{eq:lb_hermitian}).
For Hermitian systems, $\text{Im}(1/V_\alpha)=0$. In this case,
Eqs.~(\ref{eqapp:decayenh}) and (\ref{eqapp:LBveff}) (main text)
can be written as
\begin{equation}
\frac{\gamma^*}{\gamma_0} = 1 + \frac{3\pi\,
  c^3}{\omega_0^2}\,\sum_\alpha\text{Re}\left( \frac{1}{V_\alpha}\right)\text{Im}\left(\frac{1}{\omega_\alpha-\omega_0}\right)
\end{equation}
\begin{equation}
\frac{\Delta\omega}{\gamma_0}  = -
\frac{3\pi\,c^3}{2\omega_0^2}\,\sum_\alpha\text{Re}\left(
  \frac{1}{V_\alpha}\right)\text{Re}\left(\frac{1}{\omega_\alpha-\omega_0}\right)\; .
\end{equation}
We then define the Hermitian decay rate and Lamb shift associated to
the resonance $\alpha$ as
\begin{equation}
\frac{\gamma_\alpha^H}{\gamma_0} \equiv \frac{3\pi\,
  c^3}{\omega_0^2}\,\text{Re}\left(
  \frac{1}{V_\alpha}\right)\text{Im}\left(\frac{1}{\omega_\alpha-\omega_0}\right)
\label{eq:app:decayh}
\end{equation}
\begin{equation}
  \frac{\Delta\omega_\alpha^H}{\gamma_0} \equiv -
\frac{3\pi\,c^3}{2\omega_0^2}\,\text{Re}\left(
  \frac{1}{V_\alpha}\right)\text{Re}\left(\frac{1}{\omega_\alpha-\omega_0}\right)\; ,
\label{eq:app:lambh}
\end{equation}
so that the total decay rate and Lamb shift read $\gamma^* = \gamma_0 +
\sum_\alpha\gamma_\alpha^H$ and
$\Delta\omega=\sum_\alpha\Delta\omega_\alpha^H$ respectively.


For non-Hermitian systems, $\text{Im}(1/V_\alpha)\neq 0$. In this
case, Eqs.~(\ref{eqapp:decayenh}) and (\ref{eqapp:LBveff}) present an
extra term compared to the Hermitian case
\begin{multline}
\frac{\gamma^*}{\gamma_0} = 1 + \frac{3\pi\,
  c^3}{\omega_0^2}\,\sum_\alpha\text{Re}\left(
  \frac{1}{V_\alpha}\right)\text{Im}\left(\frac{1}{\omega_\alpha-\omega_0}\right)\\
+\frac{3\pi\,
  c^3}{\omega_0^2}\,\sum_\alpha\text{Im}\left(
  \frac{1}{V_\alpha}\right)\text{Re}\left(\frac{1}{\omega_\alpha-\omega_0}\right)
\end{multline}
\begin{multline}
\frac{\Delta\omega}{\gamma_0}  = -
\frac{3\pi\,c^3}{2\omega_0^2}\,\sum_\alpha\text{Re}\left(
  \frac{1}{V_\alpha}\right)\text{Re}\left(\frac{1}{\omega_\alpha-\omega_0}\right)\\
+\frac{3\pi\,c^3}{2\omega_0^2}\,\sum_\alpha\text{Im}\left(
  \frac{1}{V_\alpha}\right)\text{Im}\left(\frac{1}{\omega_\alpha-\omega_0}\right)\; .
\end{multline}
By making use of (\ref{eq:app:decayh}) and
(\ref{eq:app:lambh}), these expressions can be recast in the form
\begin{equation}
\frac{\gamma^*}{\gamma_0}  = 1+
\sum_\alpha \left \{ \frac{\gamma_\alpha^H}{\gamma_0} -
  2\frac{\Delta\omega_\alpha^H}{\gamma_0}\frac{\text{Im}(1/V_\alpha)}{\text{Re}(1/V_\alpha)}
\right \}
\end{equation}
\begin{equation}
\frac{\Delta\omega}{\gamma_0}  = \sum_\alpha
\left \{ \frac{\Delta\omega_\alpha^H}{\gamma_0} +
\frac{1}{2}\frac{\gamma_\alpha^H}{\gamma_0}\frac{\text{Im}(1/V_\alpha)}{\text{Re}(1/V_\alpha)}
\right \}
\end{equation}
which are the expressions (\ref{eq:decay:univ}) and (\ref{eq:lamb:univ}) of
the main text.


Now, we show how the expressions (\ref{eq:app:decayh}) and
(\ref{eq:app:lambh}) for $\gamma_\alpha^H$ and
$\Delta\omega_\alpha^H$ respectively can be rewritten in the form of Eqs.~(\ref{eq:gamma_hermitian})
and (\ref{eq:lb_hermitian}) (main text). First, by multiplying by the
complex conjugate, we can explicitely write (we remind that we defined $\omega_\alpha \equiv
\omega_\alpha'+\text{i}\,\omega_\alpha''$)
\begin{equation}
\text{Re}\left( \frac{1}{\omega_\alpha - \omega_0}\right) =
\frac{\omega_\alpha'-\omega_0}{|\omega_\alpha -
  \omega_0|^2}=\frac{\omega_\alpha' - \omega_0}{(\omega_\alpha' -
  \omega_0)^2 + \omega_\alpha''^2}
\end{equation}
\begin{equation}
\text{Im}\left( \frac{1}{\omega_\alpha - \omega_0}\right) =
\frac{-\omega_\alpha''}{|\omega_\alpha -
  \omega_0|^2}=\frac{-\omega_\alpha''}{(\omega_\alpha' -
  \omega_0)^2 + \omega_\alpha''^2}
\end{equation}
By reporting these expressions into Eqs.~(\ref{eq:app:decayh}) and
(\ref{eq:app:lambh}), we get
\begin{equation}
\frac{\gamma_\alpha^H}{\gamma_0} = \frac{3\pi\,
  c^3}{\omega_0^2}\,\text{Re}\left(
  \frac{1}{V_\alpha}\right)\frac{-\omega_\alpha''}{(\omega_\alpha' -
  \omega_0)^2 + \omega_\alpha''^2}
\end{equation}
\begin{equation}
  \frac{\Delta\omega_\alpha^H}{\gamma_0} = -
\frac{3\pi\,c^3}{2\omega_0^2}\,\text{Re}\left(
  \frac{1}{V_\alpha}\right)\frac{\omega_\alpha' - \omega_0}{(\omega_\alpha' -
  \omega_0)^2 + \omega_\alpha''^2}
\end{equation}
Finally, by introducing the Purcell factor defined in
Eq.~(\ref{eq:purcell_factor}) (main text), we end up with the Eqs.~(\ref{eq:gamma_hermitian})
and (\ref{eq:lb_hermitian}) of the main text, that is
\begin{equation}
\frac{\gamma_\alpha^H}{\gamma_0} = F_\alpha \left(
  \frac{\omega_\alpha'}{\omega_0} \right)^2
\frac{\omega_\alpha''^2}{(\omega_\alpha' - \omega_0)^2 +
  \omega_\alpha''^2} 
\label{eq:app:gammah_full}
\end{equation}
\begin{equation}
\frac{\Delta\omega_\alpha^H}{\gamma_0} = F_\alpha \left(
  \frac{\omega_\alpha'}{\omega_0} \right)^2 \frac{\omega_\alpha''}{2} \frac{\omega_\alpha'-\omega_0}{(\omega_\alpha' - \omega_0)^2 +
  \omega_\alpha''^2}\; .
\label{eq:app:lbh_full}
\end{equation}

\subsection{Derivation of Eqs.~(\ref{eq:gamma_-_h}), (\ref{eq:lamb_-_h}), (\ref{eq:gamma_+}) and (\ref{eq:lamb_+})}
\label{app:derivation2}
Here we derive the expressions of the decay rate and Lamb shift, in the
single-resonance approximation, for two particular detunings of the
natural QE frequency $\omega_0$ compared to the QNM resonance
frequency $\omega_\alpha'$: $\omega_0=\omega_\alpha'\mp\omega_\alpha''$, for which the Lamb shift
presents an extremum (indicated by arrows in Fig.~\ref{fig:dielectric} (b) and (d) in the
main text).


We start with the detuning
$\omega_0=\omega_\alpha'+\omega_\alpha''$. By replacing $\omega_0$
by $\omega_\alpha'+\omega_\alpha''$ in Eqs.~(\ref{eq:gamma_hermitian})
and (\ref{eq:lb_hermitian}), one gets
\begin{equation}
\frac{\gamma_\alpha^H}{\gamma_0} = \frac{1}{2}\,F_\alpha
\,\left(  \frac{\omega_\alpha'}{\omega_\alpha'+\omega_\alpha''}\right)^2=\frac{1}{2}\,F_\alpha\,\left( \frac{1}{1-\frac{1}{2Q_\alpha}}\right)^2
\end{equation}
\begin{equation}
\frac{\Delta\omega_\alpha^H}{\gamma_0} = -\frac{1}{4}\,F_\alpha
\,\left(  \frac{\omega_\alpha'}{\omega_\alpha'+\omega_\alpha''}\right)^2
=-\frac{1}{4}\,F_\alpha\,\left( \frac{1}{1-\frac{1}{2Q_\alpha}}\right)^2
\end{equation}
where we used the fact that $Q_\alpha=-\omega_\alpha'/(2\omega_\alpha'')$.
In the single-resonance approximation, Eqs.~(\ref{eq:decay:univ}) and
(\ref{eq:lamb:univ}) thus reduce to
\begin{equation}
\frac{\gamma^*}{\gamma_0}  = 1+\frac{\gamma_\alpha^H}{\gamma_0} -
  2\frac{\Delta\omega_\alpha^H}{\gamma_0}\frac{\text{Im}(1/V_\alpha)}{\text{Re}(1/V_\alpha)}
\end{equation}

\begin{equation}
\frac{\Delta\omega}{\gamma_0}  = \frac{\Delta\omega_\alpha^H}{\gamma_0} +
\frac{1}{2}\frac{\gamma_\alpha^H}{\gamma_0}\frac{\text{Im}(1/V_\alpha)}{\text{Re}(1/V_\alpha)}\; ,
\end{equation}
and by employing the previous expressions of $\gamma_\alpha^H$ and
$\Delta\omega_\alpha^H$, one gets 
\begin{equation}
\frac{\gamma^*}{\gamma_0} = 1+\frac{1}{2}\,F_\alpha\,\left(
  \frac{1}{1-\frac{1}{2Q_\alpha}}\right)^2\left[1+\frac{\text{Im}(1/V_\alpha)}{\text{Re}(1/V_\alpha)}\right]
\label{app:1}
\end{equation}
\begin{equation}
\frac{\Delta\omega^-}{\gamma_0} = -\frac{1}{4}\,F_\alpha\,\left(
  \frac{1}{1-\frac{1}{2Q_\alpha}}\right)^2\left[1-\frac{\text{Im}(1/V_\alpha)}{\text{Re}(1/V_\alpha)}\right]\;
.
\label{app:2}
\end{equation}


Similarly, for the detuning $\omega_0=\omega_\alpha'-\omega_\alpha''$,
replacing $\omega_0$ in Eqs.~(\ref{eq:gamma_hermitian})
and (\ref{eq:lb_hermitian}) yields
\begin{equation}
\frac{\gamma_\alpha^H}{\gamma_0} = \frac{1}{2}\,F_\alpha
\,\left(  \frac{\omega_\alpha'}{\omega_\alpha'-\omega_\alpha''}\right)^2
=\frac{1}{2}\,F_\alpha\,\left( \frac{1}{1+\frac{1}{2Q_\alpha}}\right)^2
\end{equation}
\begin{equation}
\frac{\Delta\omega_\alpha^H}{\gamma_0} = \frac{1}{4}\,F_\alpha
\,\left(  \frac{\omega_\alpha'}{\omega_\alpha'-\omega_\alpha''}\right)^2
=\frac{1}{4}\,F_\alpha\,\left(
  \frac{1}{1+\frac{1}{2Q_\alpha}}\right)^2\; .
\end{equation}
Then, by plugging these equations in the expressions of the decay rate
and Lamb shift in the single-resonance case as previously,
one gets
\begin{equation}
\frac{\gamma^*}{\gamma_0} = 1+\frac{1}{2}\,F_\alpha\,\left(
  \frac{1}{1+\frac{1}{2Q_\alpha}}\right)^2\left[1-\frac{\text{Im}(1/V_\alpha)}{\text{Re}(1/V_\alpha)}\right]
\label{app:3}
\end{equation}
\begin{equation}
\frac{\Delta\omega^+}{\gamma_0} = \frac{1}{4}\,F_\alpha\,\left(
  \frac{1}{1+\frac{1}{2Q_\alpha}}\right)^2\left[1+\frac{\text{Im}(1/V_\alpha)}{\text{Re}(1/V_\alpha)}\right]\;
.
\label{app:4}
\end{equation}
Note that for the Hermitian systems, the decay rate and Lamb shift
for these two particular detunings are given by Eqs.~(\ref{app:1}), (\ref{app:2}), (\ref{app:3}) and (\ref{app:4})
with $\text{Im}(1/V_\alpha)=0$, and one ends up with the
Eqs.~(\ref{eq:gamma_-_h}) and (\ref{eq:lamb_-_h}) of the main text.

\section*{ACKNOWLEDGEMENTS}

The authors thank R\'emi Colom for fruitful discussions. E.~L. would like to thank the Doctoral
School ``Physique et Sciences de la Mati\`ere'' (ED 352) for its funding.

\bibliographystyle{apsrev4-1} 
\bibliography{sample}


\end{document}